\documentclass[twocolumn,showpacs,amsmath,amssymb]{revtex4}
\usepackage{subfigure}
\usepackage{graphicx}% Include figure files

\begin{document}

\title{Preferential survival in models of complex ad hoc networks}

\author{Joseph S. Kong}
\author{Vwani P. Roychowdhury}
 \email{{jskong,vwani}@ee.ucla.edu}
 \affiliation{Department of Electrical Engineering, UCLA, Los Angeles, CA 90095, USA}
%\date{\today}

\begin{abstract}
There has been a rich interplay in recent years between (i) empirical investigations of
real world dynamic networks, (ii) analytical modeling of the microscopic mechanisms that
drive the emergence of such networks, and (iii) harnessing of these mechanisms to either
manipulate existing networks, or engineer new networks for specific tasks. We continue in
this vein, and study the deletion phenomenon in the web by following two different sets of
web-sites (each comprising more than 150,000 pages) over a one-year period. Empirical data
show that there is a significant deletion component in the underlying web networks, but
the deletion process is not uniform. This motivates us to introduce a new mechanism of
{\em preferential survival} (PS), where nodes are removed according to the
degree-dependent deletion kernel, $D(k) \propto k^{-\alpha}$, with $\alpha\geq0$. We use
the mean-field rate equation approach to study a general dynamic model driven by
Preferential Attachment (PA), Double PA (DPA), and a tunable PS (i.e., with any
$\alpha>0$), where $c$ nodes ($c<1$) are deleted per node added to the network, and verify
our predictions via large-scale simulations. One of our results shows that, unlike in the
case of uniform deletion (i.e., where $\alpha=0$), the PS kernel when coupled with the
standard PA mechanism, can lead to heavy-tailed power law networks {\em even} in the
presence of extreme turnover in the network. Moreover, a weak DPA mechanism, coupled with
PS, can help make the network even more heavy-tailed, especially in the limit when
deletion and insertion rates are almost equal, and the overall network growth is minimal.
The dynamics reported in this work can be used to design and engineer stable {\em ad hoc}
networks and explain the stability of the power law exponents observed in real-world
networks.
\end{abstract}


% Make the title.

\maketitle

\section{Introduction} %Sec I 100 words

\subsection{Motivation and Background}

The empirical study of real-world networks such as the World Wide Web,  the movie actor
collaboration network, and scientific citation network has attracted considerable
interest. Such large-scale and complex networks have been treated as physical systems, and
stochastic models based on {\em randomized mechanisms} or {\em protocols} have been
developed to  model and explain empirically observed network characteristics.
Concomitantly, several works have shown that the network dynamic models have applications
{\em beyond} merely modeling real-world systems: It has been shown that randomized
protocols can be used to design and engineer systems, with peer-to-peer networks being the
primary example \cite{adamic_pl,phenix,sarshar,sarshar:search,sarshar:multiple,gia}.
%{\bf: Joe} ...... add references to GIA, cornell group, Huberman group's work on random walk for searching on a PL network  etc......)
One of the motivations of this work is to continue such efforts aimed at discovering new
mechanisms that play an important role in organic real-world networks, and that might be
useful in designing engineered networks and protocols.

Well-known examples of such data-inspired dynamic
models, include preferential attachment (PA) and its variants \cite{simon,barabasi,yule},
%({\bf joe:} ... I added citations to Simon and Yule....)
copying \cite{kleinberg,vazquez,krapivsky}, PA with fitness \cite{huberman,huberman_tc,bianconi},
double preferential attachment of links \cite{dorogov} and the rewiring of links
\cite{albert:local}. These mechanisms, however, model the dynamics of a growing network,
where the effect of node deletion is not considered significant. Many real-world networks
experience significant rates of node deletions. For example, nodes join and depart from
peer-to-peer networks in a random and rapid manner, and movie actors end their careers,
effectively removing themselves from collaboration networks. Hence, developing a network
dynamic model for the class of {\em ad hoc} networks with a significant deletion
component is important.

Several recently proposed models have addressed the node deletion process
\cite{chung,cooper,moore,sarshar}. However, these works take an egalitarian approach in
modeling the deletion process as {\em uniform} node failure. The uniform deletion model
fails to account for the  heterogeneity of the nodes' abilities to compete for survival,
or participate for varying periods of time in a network. Interestingly, these
\emph{uniform deletion models predict} that a network's power law (PL) degree distribution, a
signature of several real-world networks such as the Web, \emph{will disappear} as the
deletion rate becomes more significant when the primary mechanism driving the network
formation is the PA rule \cite{moore,sarshar,chung}. In order to retain a heavy-tailed
distribution (i.e., networks with PL exponent, $\gamma$, being less than $3$ and closer to
$2$) the vanilla PA mechanism has to be augmented with a dominant second mechanism that
initiates new edges from the existing nodes, such as a distributed \emph{compensation}
mechanism as introduced in \cite{sarshar}, or a  Double Preferential Attachment (DPA)
mechanism (see Section~\ref{ssec:model_dpa} and \cite{chung,cooper}). It is not clear
whether organic networks with high deletion rates naturally and inherently possess such
compensatory mechanisms to retain their empirically-observed heavy-tailed distributions.
Moreover,  while in an engineered network one might be able to enforce such stabilizing
protocols (in order to retain the advantages accrued from the natural hierarchy present in
a heavy-tailed network \cite{adamic_pl,sarshar:search}), it might be too expensive to do
so or might be too difficult to enforce, and alternative mechanisms that can stabilize the
network structure might be needed.

In the absence of any empirical studies on the node removal process of real-world
networks, the simple uniform deletion model is a reasonable assumption to work with.
We, however, ask: Is it possible to empirically study the node deletion process of an
organically grown network and quantify its characteristics? We turn to the Web, which has
proved to be a treasure trove for mechanism and modeling sleuths.
%Joe: this paragraph seems disconnected  I commented out the following to make it flow better.
%For example, a number of
%works have empirically estimated and verified the linear PA kernel from different network dataset %web crawl data
%such as movie and scientific collaboration network data and crawls of the wikipedia website\cite{jeong,newman:pa,wiki}.
%Characterizing the Web as a network with reliable nodes might have been an
%accurate description in the early days of the Web. However,
Recent empirical studies of the Web suggest that the current Web environment is extremely dynamic.
For example, Ntoulas et al. found that 20\% of the web pages in their large data set is
{\em permanently} removed in just 1 month and 50\% of the web pages are deleted in 9 months
\cite{ntoulas}. Similar findings on the short lifetime of web pages are reported in
\cite{cho,gomes}. These works, while they categorically establish that deletions of nodes
is a significant event and should be included in any dynamic modeling of the web, they do
not answer the nature of the deletion dynamics, and whether it is uniform or not.

\subsection{Summary of Results}

In a competitive network such as the Web, we expect the nodes to compete for survival in
addition to competing for links. A webpage's degree is a good approximation to its ability
to compete, since heavily linked Web documents are entitled to numerous benefits, such as
being possibly ranked higher in search engine results, and attracting higher traffic and,
thus, higher revenue through online advertisements. As a result, \emph{we conjecture} the
mechanism of {\em preferential survival} (PS), whereby each node's chance of survival
increases with its degree; in other words, pages with higher degrees would be less likely
to be deleted than their counter parts with lower degrees.

In order to verify our conjecture, we made a longitudinal study of Web data, where we
followed two different sets of web-sites (each comprising more than $150,000$ pages) over
a period of one year, as described in Sec.~\ref{sec:empirical} \footnote{Since we are
following a fixed number of sites or hosts in our crawl data, only links among the pages
in these sites are considered for our empirical analysis; links to and from these pages to
pages outside of this universe are ignored.}. We found that {\em indeed} there exists a
significant rate of node deletion in the crawl data we studied. The deletion rates $c$
(the average number of nodes deleted \emph{per} node that is added to the network, i.e.,
the network grows at the rate of $(1-c)$) for the sites we tracked  are observed to be as
high as $0.9$ (see the Appendix for further details). We next developed a method to
quantify the node deletion kernel, and found that the conjectured \emph{PS mechanism is
indeed in play} and that the degree-dependent deletion kernel (i.e., the probability that
a node of degree $k$ is deleted at any time step) behaves as, $D(k) \propto k^{-\alpha}$
($\alpha
>0$), where $\alpha$ is estimated to be 1.0 for our crawl data. Interestingly, \emph{given the high rate of node
deletion} rates in our crawl data, we \emph{found no sign} of the \emph{disappearance} of
the power law degree distribution.  %In the Appendix, we report additional
%empirical findings on estimating the deletion rate, the weak mechanism of the double
%preferential attachment of links, and measurements on the attachment kernel for our crawl data.

The empirical findings {\em motivated}   us to study the role of preferential survival
mechanism in the  well-studied stochastic PA and DPA models; see Sec. \ref{sec:model}. That
is, at every time step, in addition to adding a new node that initiates preferential edges,
an existing node is chosen according to the PS deletion kernel, $D(k) \propto k^{-\alpha}$
($\alpha >0$), and this node (along with all of its edges) is then deleted with
probability $c$. Thus, for  $\alpha=0$ it reduces to the already studied case of uniform
deletion where nodes are deleted at the rate of $c$. Otherwise, as $\alpha$ increases, the
dynamic shields higher degree nodes against deletion, \emph{even though} the overall
deletion rate remains fixed at $c$.  The main predictions of our analysis can be
summarized as follows (all analytical results are verified by large-scale simulations):
\begin{enumerate}
\item In the special case of PA and \emph{only PS} with $\alpha = 1$, our analysis shows
that the power law exponent is expected to be $\gamma=3$ for \emph{any} turnover rate $c$
between 0 and 1. Our large-scale simulation results are in good agreement with our
analysis. Thus, the \emph{ PS mechanism by itself can arrest the divergence of the PL
exponent predicted for the uniform deletion case} (i.e., $\alpha=0$).
Furthermore, we analytically derive the node lifetime distribution for the preceding case of PS
(with $\alpha=1.0$) and PA, and find that the
probability a given node survives for $t$ time steps, converges to a constant as $t$
grows.  The analytical distribution closely matches the empirical distribution of
lifetimes in our crawl, providing further credence to the model.

\item As a comparison, when we study the case of uniform deletion and DPA, we find
 that in order for the PL exponent to be stabilized at around $3$ for high rates of deletions (i.e., when
$c\approx 1$), the number of doubly preferential (DP) edges have to be increased
significantly, i.e., if at every time step, each incoming node brings in $m$ edges on the
average, then the existing nodes have to initiate $bm$ DP edges at every time step, where
$b>>1$. Thus, one needs a very strong DPA component to compensate for a uniform deletion
case.

\item In the case of \emph{both} PS and DPA,  we show that the power law exponent actually
 \emph{decreases} as the network experiences higher turnover rate.  Thus, when used in
 conjunction with the PS dynamic, even a weak DPA mechanism (i.e. for example for the empirically
 estimated values of $m=10$ and $b=0.1$) can be critical in driving the power law exponent close to 2 even
 in the face of extremely high rate of turnovers.
\end{enumerate}
Although the PS mechanism is inspired by empirical Web dynamics, a complete model of the
Web should take into account other factors such as the nodes' varying \emph{fitness} in
attracting links \cite{huberman_tc,bianconi}, and such a modeling effort is beyond the
scope of this paper. Moreover, while deletion is dominated by the PS mechanism for the two
sets of crawls studied in this paper, further work studying a larger web sample is needed
to justify a general conclusion that PS is a dominant mechanism for all parts or a
majority of the web.

However, as our  analysis and simulations indicate, the
 PS dynamic when incorporated into the PA and DPA models leads to a stable structure and can
be used to model and design protocols to engineer large scale complex networks.
Potential implications of the PS mechanism for both modeling and designing real-world
network application purposes are discussed further in Section~\ref{sec:application}.

\section{Empirical Measurements}\label{sec:empirical}
\subsection{The Dataset}%and Resilience of the Power Law} %Sec II 100 words
Our dataset of the World Wide Web was obtained from the Stanford WebBase project 
\footnote{http://dbpubs.stanford.edu:8091/\~{}testbed/doc2/WebBase/}. We
randomly selected a set of web hosts, comprising of roughly 170 thousand pages, and
tracked their evolution monthly for the year 2006. This dataset is denoted as SET1. In
order to further validate our results, we sampled another set of web hosts comprising of
roughly 150 thousand pages and tracked their evolution for the same period. For this
dataset, more than 99\% of the nodes belong to the weakly  giant connected component
(i.e., when we look at all edges as undirected) over the examined period. This dataset is
denoted as SET2. The nodes in The WebBase crawler would extract a maximum of 10 thousand
pages per host. However, the 10 thousand pages per host limit is not a problem since none
of the tracked hosts reaches this limit.
\begin{figure}[htp]
\subfigure[\small ]{
\centering
\includegraphics[height=2.3in,angle=-90]{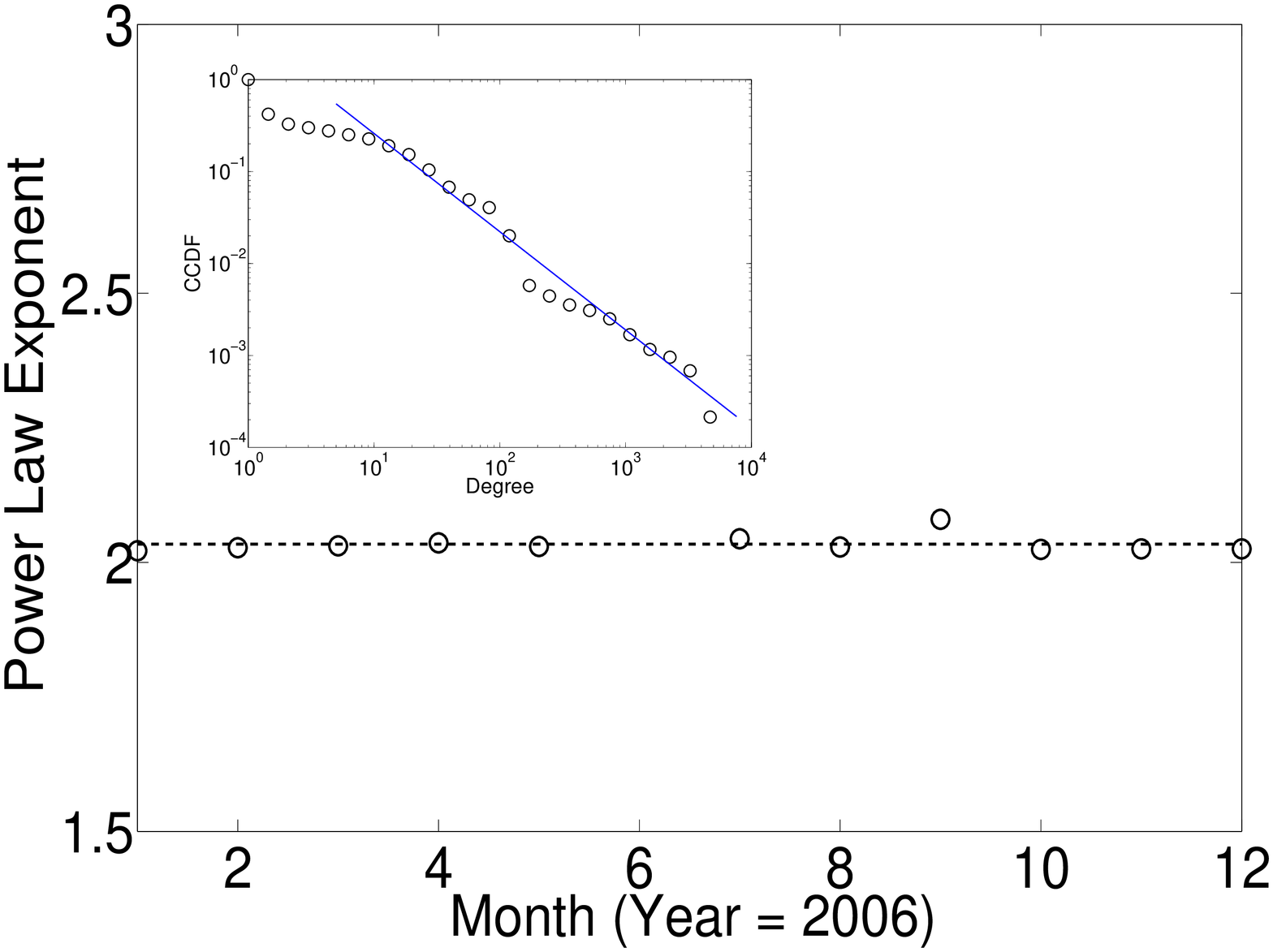}
\label{fig_2}
}
\subfigure[\small ]{
\centering
\includegraphics[height=2.3in,angle=-90]{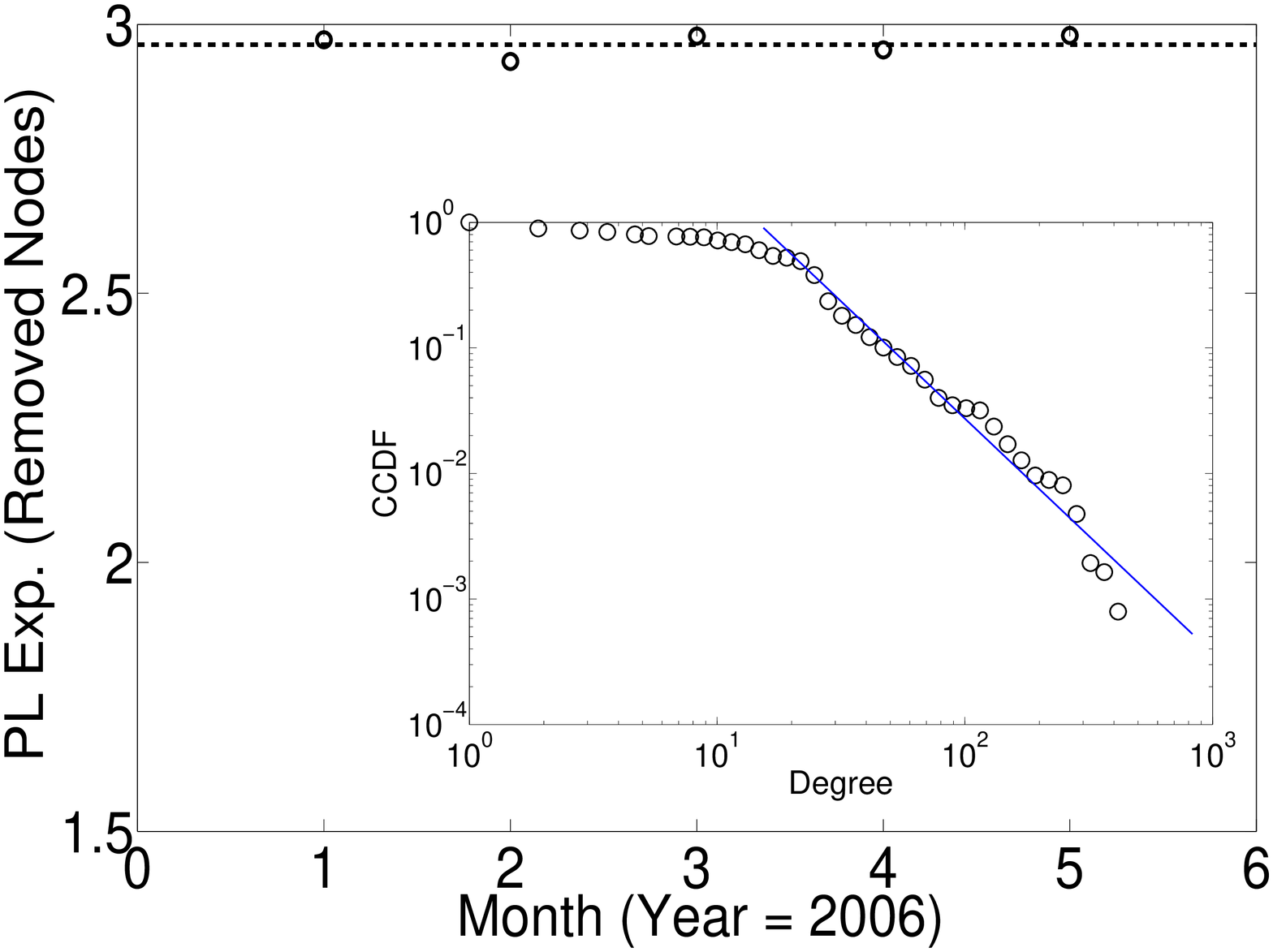}
\label{fig_3}
}
\subfigure[\small ]{
\centering
\includegraphics[width=2.3in]{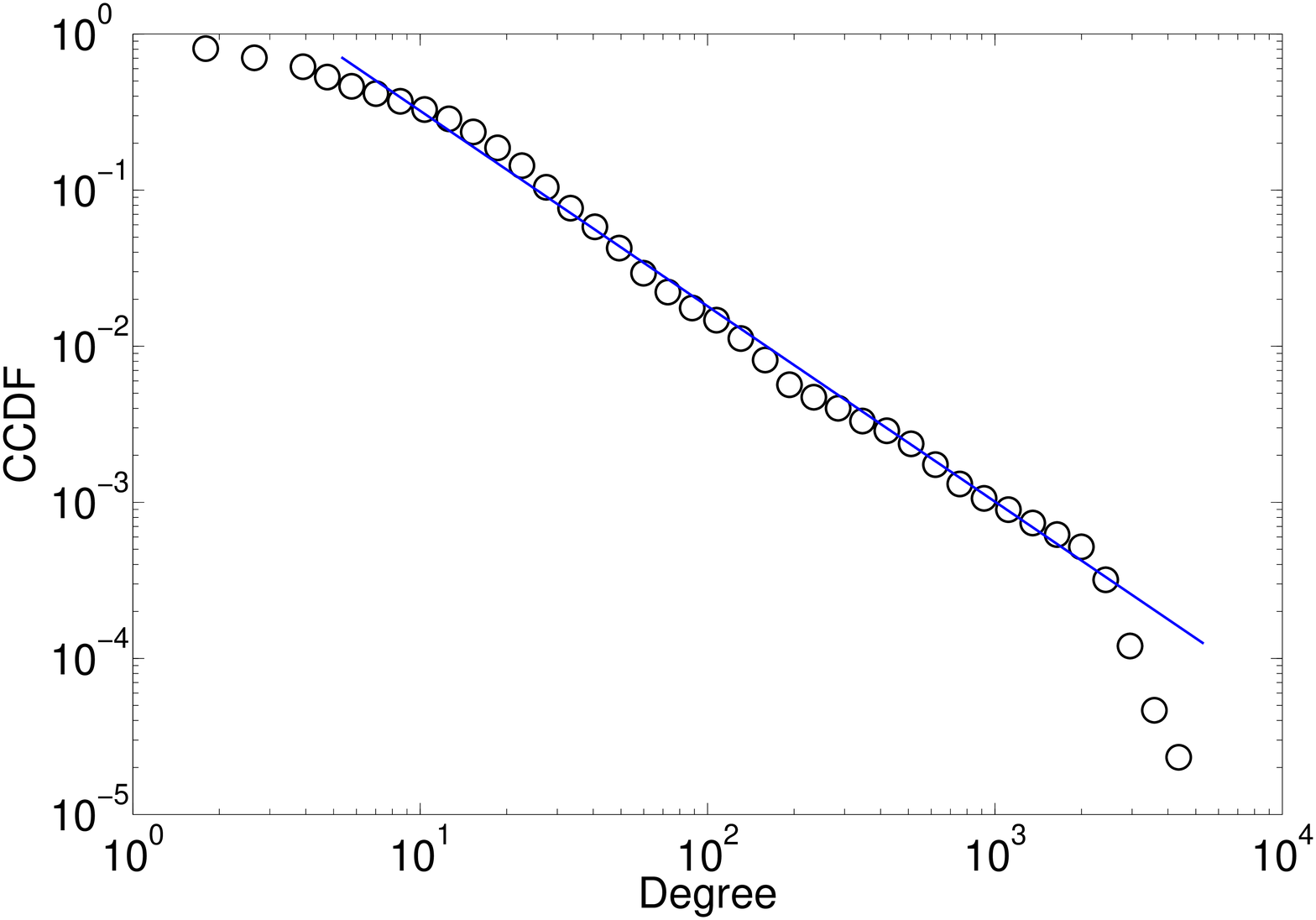}
\label{gccfig}
}
\subfigure[\small ]{
\centering
\includegraphics[width=2.3in]{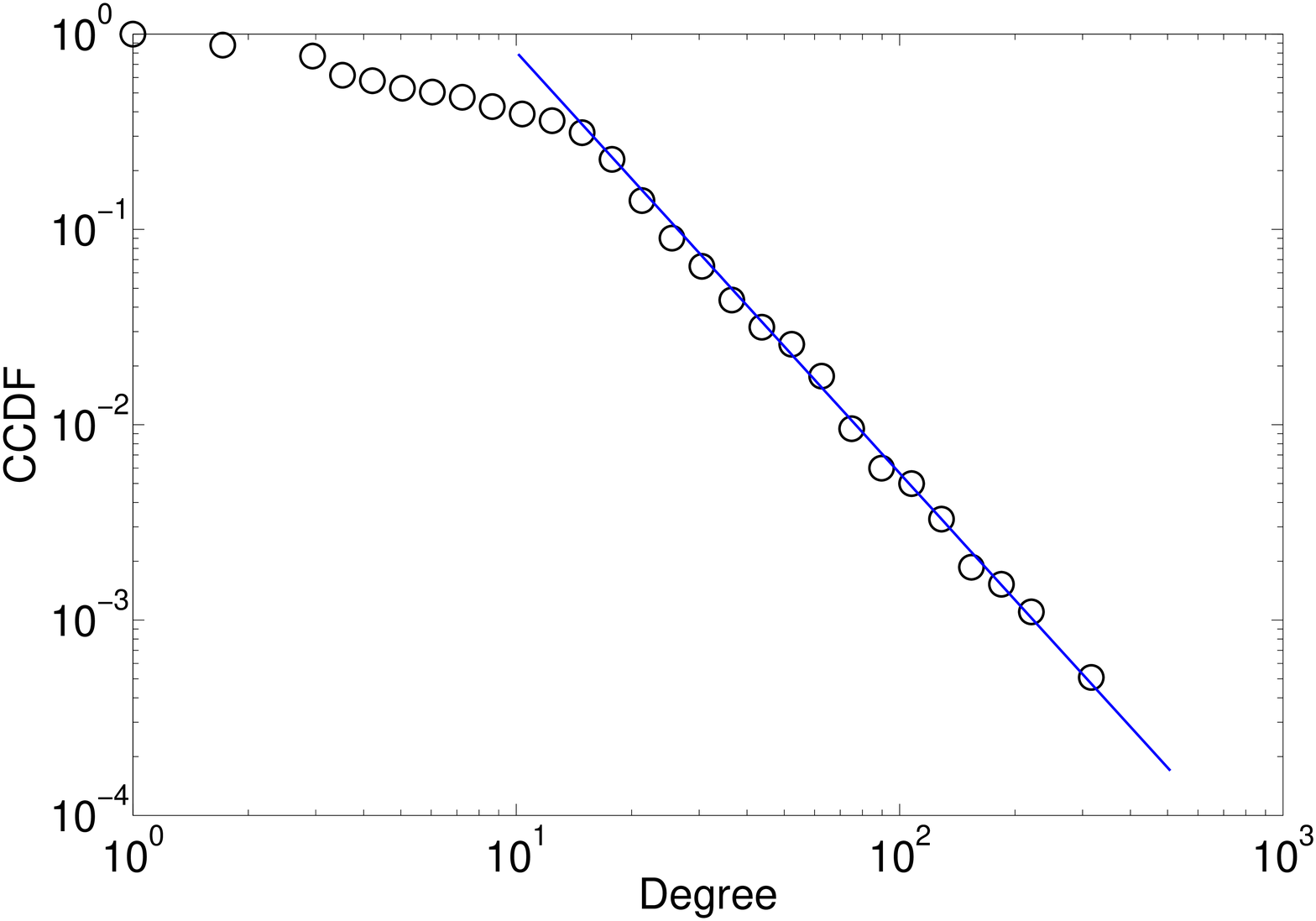}
\label{gccfig_rem}
}
\caption{\small (a) SET1: The power law exponent of the degree distribution of the
sampled Web for each month in 2006. The inset figure shows the degree
distribution for May, 2006.
%Note that the CCDF (complementary cumulative distribution function) is plotted.
(b) SET1: the power law exponent of the degree distribution of the set of removed nodes for different months in 2006.
The inset figure shows the power law degree distribution for the set of webpages that are removed
in March, 2006. %For February 2006, we identify all new edges emerging among nodes that already exist in January 2006 and record the degrees of the target nodes of these new edges.  We plot the number of edges normalized by the joint degree distribution versus the degree product.
(c) SET2: the degree distribution for June, 2006.  The power law exponent is $\gamma = 2.2$.
(d) SET2: the power law degree distribution for the set of webpages that are removed in June, 2006.
The power law exponent is $\gamma_{del} = 3.2$.
}
\end{figure}

\subsection{Evidence of Preferential Survival of Webpages} %100 words
We mined our Web dataset for direct empirical evidence of the preferential survival mechanism.
We regard the web graph as an undirected network and investigate the degree distribution
of the set of deleted nodes in a given month (i.e. the set of nodes that are alive in a given month but disappear in the
following month). If nodes were to be deleted uniformly randomly, the degree distribution
of the set of deleted nodes would be {\em identical} to the network's degree distribution.
For SET1, we found that the power law exponent of the degree distribution of the set of deleted
nodes to be $\gamma_{del} \approx 3.0$ (Fig. \ref{fig_3}), which is significantly {\em
different} from the power law exponent for the entire network $\gamma \approx 2.0$ (Fig. \ref{fig_2}).
We found similar results for SET2: the power law exponent of the degree distribution of the set of removed
nodes to be $\gamma_{del} = 3.2$ (Fig. \ref{gccfig_rem}), which is significantly greater
than the power law exponent for the entire network $\gamma = 2.2$ (Fig. \ref{gccfig}).

Our finding from both SET1 and SET2 suggests that a node is removed according to the {\em deletion probability
kernel}: $D(k) \propto k^{-\alpha}$, where $\alpha = \gamma_{del} - \gamma \approx 1.0$ in
our case. We will show in our model (see Sec. \ref{ssec:model_ps}) that a deletion kernel with $\alpha=1$ leads to the {\em stabilization} of
the power law exponent at $\gamma=3$, for any turnover rate $c$ between 0 and 1. \iffalse
This is in stark contrast of the uniform deletion model, where high turnover rate leads to
a diverging power law exponent.\fi

\subsection{Resilience of the Power Law Exponent}
We tracked the PL exponent, $\gamma$, for our Web dataset with very high turnover rates (see Appendix).
The power law exponent does not show any sign of divergence and is highly resilient under high rate of turnover.
For SET1, the exponent stays around $\gamma \approx 2.0$ (Fig. \ref{fig_2}); to be self-consistent,
{\em only edges linking the tracked pages} are considered in estimating the degree distributions.
For SET2, the exponent is around $\gamma \approx 2.2$ over the examined period.

\section{The Model}\label{sec:model} %Sec V 100 words
In order to study the implication of the preferential survival mechanism,
we propose the following dynamic model: at each time step, a node
joins the network and makes $m$ links to $m$ nodes preferentially; with probability $c$, a
node is chosen to be removed, according to the deletion kernel $D(k) \propto k^{-\alpha}$,
along with all of its associated links; $bm$ {\em new} internal edges link in a double
preferential attachment (DPA) manner to existing nodes.
The parameter $c$ denotes the turnover rate or the deletion rate, which is defined as the rate of node removal
divided by the rate of node addition.

Each node in the network is labeled by its insertion time.  Let $D(i,t)$ be the probability that the $i$th node
is still in the network at time $t$, where $t>i$. Note that $D(i,t)$ yields the {\em lifetime distribution} of node $i$.
We have:
\begin{equation}
D(i,t+1) = D(i,t)[1-c\frac{k(i,t)^{-\alpha}}{N(t)\langle k^{-\alpha}(t)\rangle }]. \label{D_eq}
\end{equation}
The initial condition is $D(i,i)=1$ and $\langle k^{-\alpha}(t)\rangle = \sum_{k}
k^{-\alpha}P(k,t)$, which can be considered as the "-$\alpha$" moment of the degree distribution at time $t$
(see Table \ref{tab:var_def} for the definition of symbols).

Assuming the $i$th node is still in the network at time $t$,
the evolution of its expected degree is described by the following equation:
\begin{eqnarray}\label{eq:rate_1}
    \frac{\partial k(i,t)}{\partial t}&=&m\frac{k(i,t)}{S(t)}-ck(i,t)P(\hbox{a neighbor is removed}) \nonumber\\
                                      &+&2b m\frac{k(i,t)}{S(t)},
\end{eqnarray}
where the sum of node degrees at time $t$ is described by $S(t)=\langle k(t)\rangle N(t)$,
with $\langle k(t)\rangle$ denoting the average node degree at time $t$ and $N(t)=(1-c)t$ is the number of nodes at time $t$.

\begin{table}[t]
\centering
\begin{tabular}{|c | c |}
\hline {\scriptsize\textbf{Var.}}         & {\scriptsize \textbf{Definition}} \\ \hline
{\scriptsize $k(i,t)$}             & {\scriptsize expected degree of the $i$th node at
time $t$}   \\ \hline {\scriptsize $S(t)$}               & {\scriptsize sum of node
degrees at time $t$}   \\ \hline {\scriptsize $N(t)$}               & {\scriptsize size of
the network at time $t$}   \\ \hline
{\scriptsize $\langle k(t)\rangle$}   & {\scriptsize average node degree at time $t$} \\ \hline
{\scriptsize $\langle k_{del}(t)\rangle$}   & {\scriptsize average degree of  a deleted node at time $t$} \\ \hline
{\scriptsize $\langle k^{-\alpha}(t)\rangle$} & {\scriptsize $\sum_{k} k^{-\alpha} P(k,t)$} \\ \hline
{\scriptsize\textbf{Const.}} & {\scriptsize \textbf{Definition}} \\ \hline
{\scriptsize $m$} & {\scriptsize number of connections of the joining node}   \\ \hline
{\scriptsize $c$} & {\scriptsize turnover rate or number of nodes deleted in each time step}   \\ \hline
{\scriptsize $b$} & {\scriptsize ratio of number of internal edges added per time step} \\
                  & {\scriptsize and number of connections per joining node}   \\ \hline
{\scriptsize $\alpha$} & {\scriptsize exponent in the deletion kernel $D(k) \propto k^{-\alpha}$}   \\ \hline
{\scriptsize $a_0$} & {\scriptsize the "-1" moment of the degree distribution: $\sum_{k} k^{-1} P(k)$}   \\ \hline
\end{tabular}
\caption{Table of Definitions}
\label{tab:var_def}
\end{table}

\begin{figure*}[htp]
\subfigure[\small ]{
\centering
\includegraphics[width=3.1in]{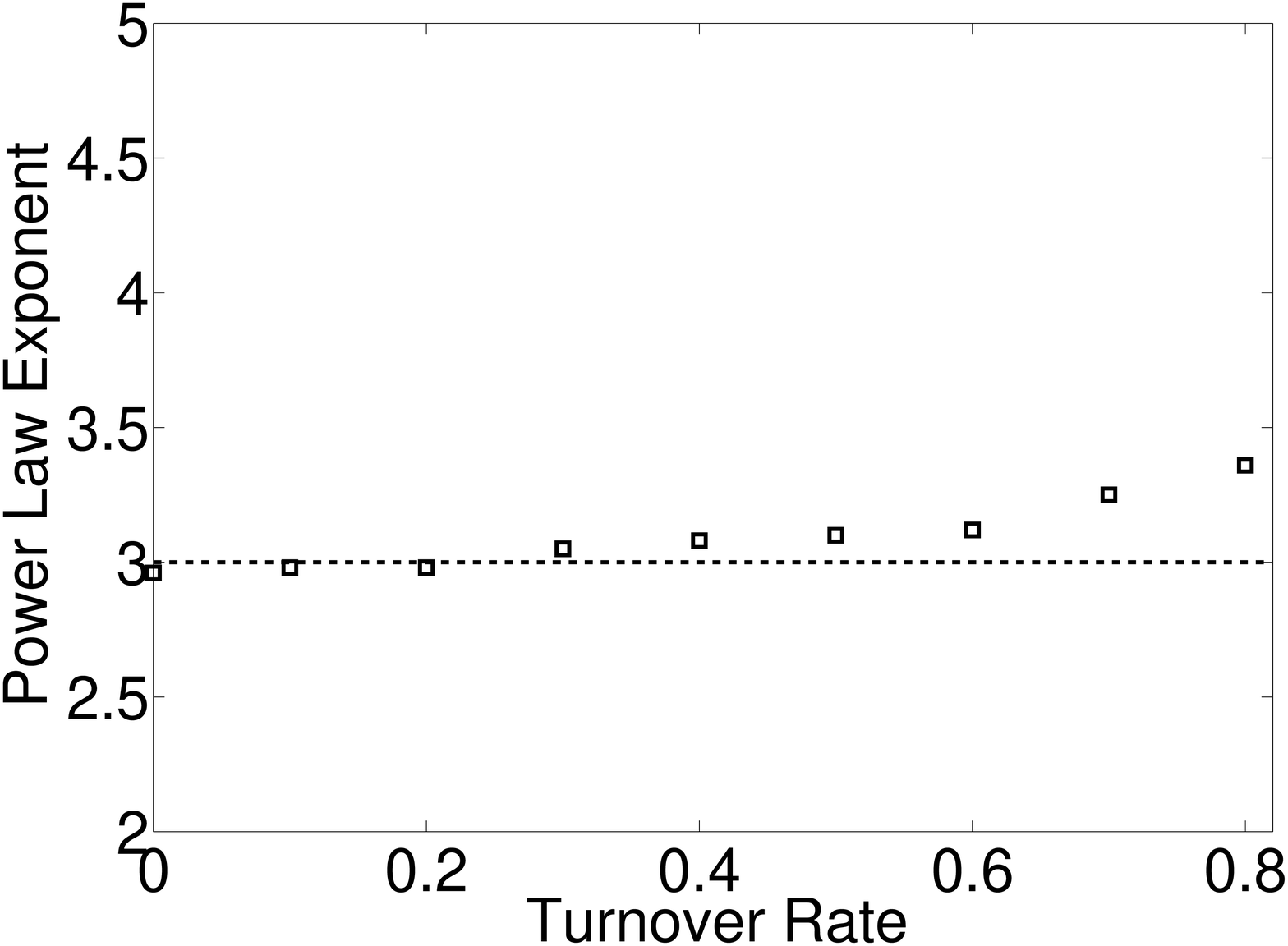}
\label{supfig_3}
}
\subfigure[\small ]{
\centering
\includegraphics[width=3.1in]{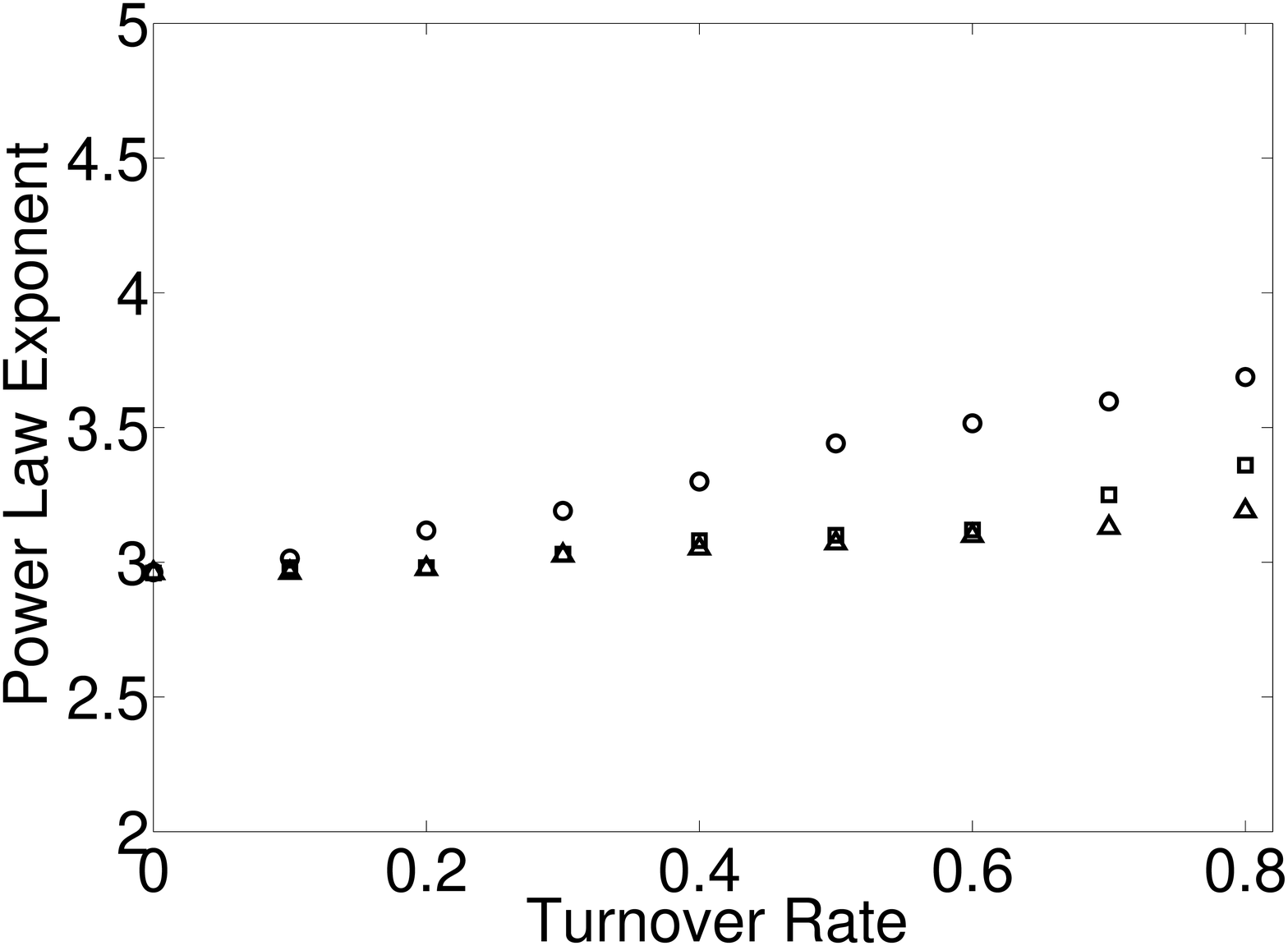}
\label{supfig_all}
}
\caption{\small Power law exponent for the degree distribution of networks generated by simulation ($m=8$, $b=0$).
At the time the snapshots are taken, the networks reach 20 000 nodes.
(a) Preferential survival $\alpha=1.0$ (squares): the points do not deviate from 3 for $c\leq0.6$ (see Eq. (\ref{eq1:gamma})).
For $c > 0.6$, the simulation points deviate from $\gamma = 3$ slightly due to finite number of time steps in simulations.
(b) Preferential survival $\alpha=0.5$ (circles), $\alpha=1.0$ (squares), $\alpha=1.5$ (triangles):
simulations results indicate that a greater $\alpha$ slows down the increase of the power law exponent.}
\end{figure*}

The initial condition is: $k(i,i)=m$.  Eq. (\ref{eq:rate_1}) gives the rate at which the $i$th node gains connections at time $t$.
%The preferential attachment kernel is given by $\frac{k(i,t)}{S(t)}$.
The first term in Eq. (\ref{eq:rate_1}) describes the attachments of the $m$ preferential links as a result of the joining node;
the second term denotes the deletion of node $i$'s neighbors according to the deletion kernel;
the third term describes the appearance of $b m$ new internal
edges attaching in a double preferential manner to $2b m$ target nodes.  Furthermore, the evolution of $S(t)$ is described by:
\begin{equation}\label{eq:rate_3}
    \frac{\partial S(t)}{\partial t} = 2(1+b)m - 2c\langle k_{del}(t)\rangle
\end{equation}
where $\langle k_{del}(t)\rangle$ is the average degree of a deleted node at time $t$.

Eq. (\ref{eq:rate_3}) gives the rate of increase for the sum of node degrees at time $t$;
the first term on the right hand side describes the addition of $(1+b)m$ edges, hence $2(1+b)m$ degrees
are added to the sum of degrees; the second term describes the loss of edges as a result of the removed node.

Now to calculate the power-law exponent, we note that
\begin{eqnarray}
    P(k,t)&=& \frac{\hbox{No. of nodes with degree $=\ k$}}{\hbox{Total number of nodes}} \nonumber \\
          &=& \frac{1}{N(t)}\sum_{i:k(i,t)=k} D(i,t)\nonumber \\
          &=&\displaystyle \frac{1}{N(t)}D(i,t)\left\vert\frac{\partial k(i,t)}{\partial i}\right\vert_{i:k(i,t)=k}^{-1} \label{P_eq}
\end{eqnarray}

The general model stated above appears to be very difficult to solve analytically.
%and hence, we first study special cases, before presenting numerical solutions to the general case.

\subsection{PA with PS ($\alpha>0$)}\label{ssec:model_ps}
We now consider the preferential survival model by setting the parameters $\alpha=1.0$ and $b = 0$,
where the value of $\alpha$ is inspired by empirical measurements of the Web.  We first note that
\begin{equation}
P(\hbox{ a node of degree } k \hbox{ is deleted}) = \frac{k^{-1}}{N(t)a_0}N(t)P(k,t)
\end{equation}
where $a_0=\sum_k k^{-1}P(k)$, under the assumption that $a_0(t)=\sum_k k^{-1}P(k,t)$ converges rapidly to the stationary
value $a_0$.  This assumption has been verified numerically.
We thus obtain that $\langle k_{del}(t)\rangle = 1/a_0$.
It is simple to show that the evolution of the sum of degrees at time $t$ is given as:
\begin{equation}\label{eq1:rate_3}
    \frac{\partial S(t)}{\partial t} = 2m - 2\frac{c}{a_0}
\end{equation}
 We now obtain: $S(t) = (2m - 2c/a_0)t$.

Similarly, we invoke the assumption that $\langle k(t)\rangle =\sum_k kP(k,t)$ converges quickly to the stationary constant $\langle k\rangle$ and
verified this assumption numerically.  Now, assuming the $i$th node's neighbors have the average degree $\langle k\rangle$,
the evolution of the expected degree of the $i$th node at time $t$ is described by the following equation after performing some calculations:
\begin{equation}
\frac{\partial k(i,t)}{\partial t} = m\frac{k(i,t)}{S(t)}-ck(i,t)\frac{\langle k \rangle^{-1}}{N(t)a_0} = \frac{k(i,t)}{2t}
\end{equation}
The equation above implies that
\begin{equation}
k(i,t) = m(\frac{t}{i})^{\beta} \label{k_eq}
\end{equation}
where $\beta=1/2$.

After substituting Eq. (\ref{k_eq}) into Eq. (\ref{D_eq}), one can show that the $D(i,t)$ equation is described by the following:
\begin{equation}
D(i,t) = \frac{1}{e^{2c_0}} e^{2c_0(t/i)^{-1/2}} \label{eq:life}
\end{equation}
where $c_0 = \frac{c}{(1-c)ma_0}$. It is interesting to note that $D(i,t)$ has an initial exponential decay and
converges to a positive constant as $t \rightarrow \infty$.

We now invoke Eq. (\ref{P_eq}) to get the stationary degree distribution:
\begin{eqnarray}
    P(k)&=& 2\frac{m^2e^{-c_0}}{1-c}e^{c_0(k/m)^{-1}}k^{-3} \label{eq1:gamma}
\end{eqnarray}
which has a power law tail with the exponent $\gamma = 3$.  This is the same power law exponent obtained for the
simple preferential attachment model with no deletions.  The analytical result is verified by
large-scale simulations (see Fig. \ref{supfig_3}).  Thus, {\em we found
that preferential survival is the self-stabilization mechanism that nulls the harmful effect node deletion has on the power law exponent}.
Consequently, the power law exponent remains at 3 even in the face of node turnovers under the preferential survival mechanism.

For the case of non-unity $\alpha$, we resort to simulation studies:
for $\alpha=0.5$ the divergence observed for the uniform deletion case is checked, i.e.,
the PL exponent does not go to infinity as $c$ approaches 1, but still can be much larger
than $3$; for $\alpha = 1.5$ (i.e., the high-degree nodes are now being protected even more),
the distribution becomes slightly more heavy tailed (see Fig. \ref{supfig_all}).

{\bf Lifetime Distribution of Webpages.}
In addition to the degree distribution, we study the lifetime characteristics of webpages.
In order to obtain the empirical lifetime distribution of webpages,
we gathered and processed additional crawls from the period 2003 to 2005 from WebBase (SET1).
Our analytical model predicts that the probability a given node survives for $l$
time steps, has an initial exponential decay, followed by a slow convergence to a positive
non-zero constant as $l$ grows: $P(l) \propto e^{c_0/l^\tau}$, where $c_0$ and $\tau$ are
positive constants (see Eq. (\ref{eq:life})). In other words, a given node has a non-zero probability  of
achieving a very long, if not an eternal, life.

\begin{figure}[htp]
\centering
\includegraphics[width=3.1in]{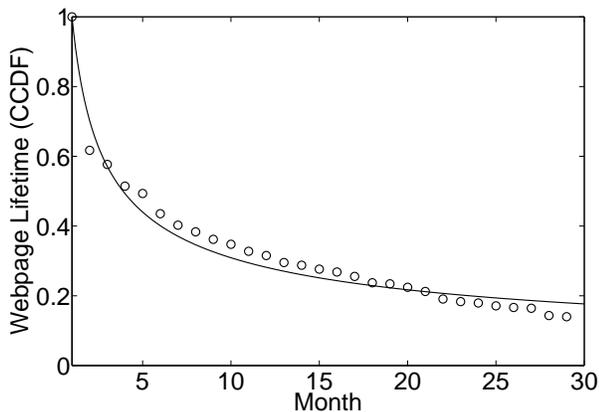}
\caption{\small The figure plots the lifetime distribution of webpages from our sampled Web. The empirical distribution matches well with
the analytical function predicted by our model.}
\label{lifetime}
\end{figure}

Our model, thus, offers an explanation to
the empirical observation that a significant fraction of webpages has short lifetimes,
while some webpages persist for a very long time (Fig. \ref{lifetime}).
Empirical webpage lifetime distribution of similar form has also been obtained by another measurement study \cite{ntoulas},
but no theoretical explanation has been offered.
%Our preferential survival based model for predicting the lifetime distribution of webpages
%enables developers to more efficiently design and manage data structures and algorithms
%for Web applications, and has been long sought for.

\subsection{PA and DPA with Uniform Deletion ($\alpha=0$)}\label{ssec:model_dpa}
We now consider the double preferential attachment (DPA) with uniform deletion model by
setting the parameter $\alpha = 0$. Using different methods, the same model has been
analyzed in \cite{chung,cooper}. The evolution of the expected degree of the node born in
time step $i$ at time $t$ is described by the following equation:
\begin{equation}\label{eq0:rate_1}
    \frac{\partial k(i,t)}{\partial t}=m\frac{k(i,t)}{S(t)}-c\frac{k(i,t)}{N(t)}+2b m\frac{k(i,t)}{S(t)},
\end{equation}
where $N(t)=(1-c)t$, and $S(t)$ is described by
\begin{equation}\label{eq0:rate_3}
    \frac{\partial S(t)}{\partial t} = 2(1+b)m - 2c\frac{S(t)}{N(t)} = 2(1+b)m - 2c\frac{S(t)}{(1-c)t}
\end{equation}

%When we have $\alpha = 0$, nodes are deleted uniformly randomly.  Hence, the probability that a node's given
%neighbor is selected for deletion is $1/N(t)$.  The average degree of a deleted node is given by $S(t)/N(t)$.

Solving Eq. (\ref{eq0:rate_3}), we get:
\begin{equation}\label{eq0:rate_4}
S(t) = 2(1+b)m\frac{1-c}{1+c}t
\end{equation}

Now, Eq. (\ref{eq0:rate_1}) becomes:
\begin{eqnarray}
\frac{\partial k(i,t)}{\partial t} &=& \frac{k(i,t)}{2(1-c)t}(\frac{1-c+2b}{1+b}) \label{eq0:rate_5}
\end{eqnarray}

Solving Eq. (\ref{eq0:rate_5}), we get:
\begin{equation}\label{eq0:rate_6}
    k(i,t) = m(\frac{t}{i})^{\beta}\ ,
\end{equation}
where $\beta = \frac{1-c+2b}{2(1-c)(1+b)}$.

In addition, we solve the $D(i,t)$ equation and obtain:
\begin{equation}
D(i,t) = (\frac{t}{i})^{c/(c-1)}
\end{equation}

After invoking Eq. (\ref{P_eq}), the scaling relation for the power law exponent $\gamma$ can be derived as:
\begin{equation}
\gamma = 1 + \frac{1}{(1-c)\beta} = 1 + \frac{2(1+b)}{1-c+2b} \label{eq0:rate_8}
\end{equation}
The analytical prediction matches closely with large-scale simulation results as shown in Fig. \ref{supfig_2}.
From our Web dataset, the parameter $b$ is estimated to be quite low at $b \approx 0.1$ (see Appendix).  For such low value of $b$, the
power law exponent is expected to diverge rapidly as the turnover rate $c$ moves away from zero as shown in Eq. (\ref{eq0:rate_8}).
Thus, the emergence of the double preferential attachment edges is not sufficient to explain the observed resilience
of the power law under high rate of turnover.

\begin{figure*}[htp]
\subfigure[\small ]{
\centering
\includegraphics[width=3.1in]{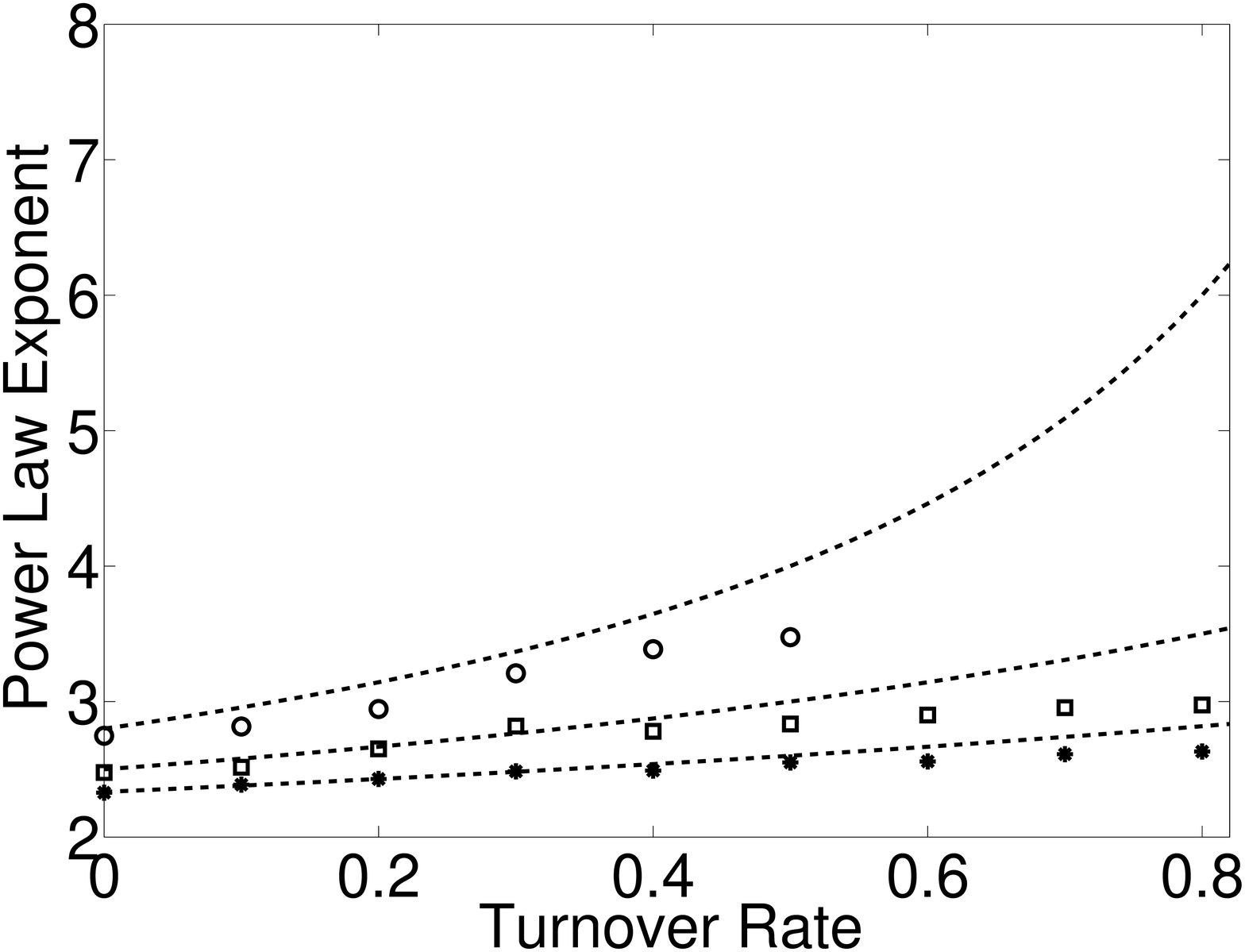}
\label{supfig_2}
}
\subfigure[\small ]{
\centering
\includegraphics[width=3.1in]{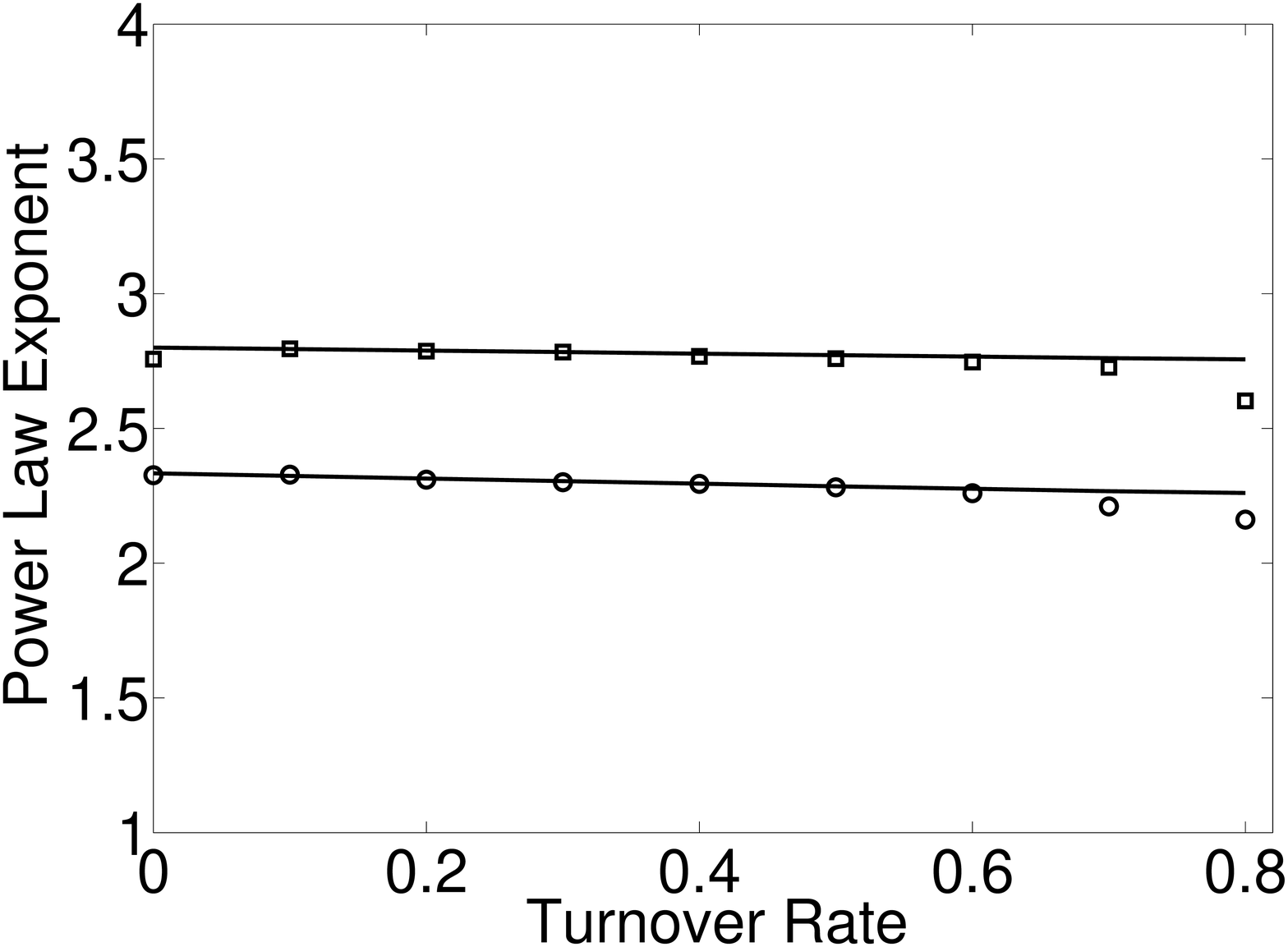}
\label{supfig_4}
}
\caption{\small Power law exponent for the degree distribution of networks generated by simulation ($m=8$).
At the time the snapshots are taken, the networks reach 20 000 nodes.
Double Preferential Attachment (DPA): Power law exponent for the degree distribution of networks generated with
analytical models for uniform deletion ($m=8$, $\alpha=0$): $b=1/8$ (circles), $b=1/2$
(squares), $b=1$ (stars). Note that tracking high power law exponent values much above $3.0$ is rather
difficult, since the distribution is rapidly decreasing and the power law region is
typically exhibited over less than one decade. Hence, some data points for $b=1/8$ are
omitted. %The simulation points show a good fit with the theoretical prediction from Eq. (\ref{eq:rate_8}) (dashed lines).
As shown, unless $b$ is unreasonably high, DPA alone cannot stop the divergence of the PL exponent under heavy turnover.
(b) Preferential survival with double preferential attachment edges ($\alpha=1$): $b=1/8$ (squares) and $b=1$ (circles).
The simulation results are in agreement with the theoretical prediction from Eq. (\ref{eq2:gamma}). }
\end{figure*}

\subsection{PA and DPA with PS}\label{ssec:model_psdpa}
We now investigate the preferential survival with double preferential attachment model by setting the parameters $\alpha=1.0$ and for general $b$.
It is simple to show that the evolution of $S(t)$ is described by the following equation:
\begin{equation}\label{eq2:rate_3}
    S(t) = (2m(1+b) - 2\frac{c}{a_0})t
\end{equation}

As in the previous section, the assumption of the fast convergence of $a_0$ and $\langle k\rangle$ is numerically verified and used.
The evolution of the expected degree of the $i$th node at time $t$ is described by the following:
\begin{equation}
\frac{\partial k(i,t)}{\partial t} = m\frac{k(i,t)}{S(t)}-ck(i,t)\frac{\langle k\rangle^{-1}}{N(t)a_0} + 2bm\frac{k(i,t)}{S(t)}
\end{equation}
Solving the above equation, we get:
\begin{equation}
k(i,t) = m(\frac{t}{i})^{\beta} \label{eq2:k_eq}
\end{equation}
with
\begin{equation}
\beta = \frac{1}{2}(1+\frac{bm}{(1+b)m-c/a_0})  \label{eq2:beta}
\end{equation}
where $\langle k\rangle$ is the average degree in the network.  Finally, the power law exponent is given by:
\begin{equation}
\gamma = 1+\frac{1}{\beta} = 1+(\frac{1}{2}(1+\frac{bm}{(1+b)m-c/a_0}))^{-1} \label{eq2:gamma}
\end{equation}
Note that Eq. (\ref{eq2:beta}) is a strictly increasing function for $0 \leq c \leq 1$,
assuming that $a_0$ stays roughly constant for different value of $c$ (this assumption has been numerically verified).
Thus, the power law exponent actually {\em decreases} as the turnover rate increases.
After obtaining the value of $a_0$ numerically, the analytical
results are verified by large-scale simulations (see Fig. \ref{supfig_4}).

In summary, our model predicts the following: for $\alpha=0$, our model reduces to the
case of uniform deletion, where the power law exponent is predicted to diverge for
moderate amount of DPA edges. In the case of PS with $\alpha = 1$, our analysis shows that
the power law exponent is expected to be $\gamma=3$ for \emph{any} turnover rate $c$
between 0 and 1.
%Our large-scale simulation results are in good agreement with our analysis (Fig. \ref{fig_4}).
Thus, the preferential survival mechanism by itself can prevent the divergence of the
PL exponent predicted for the uniform failure case. \iffalse  In other words, it
reinforces and stabilizes the network's degree hierarchy (i.e. the power law degree
distribution). ; otherwise, the natural hierarchy  vanishes under uniform deletion. \fi
Furthermore, PS aided by a  weak DPA dynamic can reinforce and stabilize the network's
degree hierarchy even more as $c$ approaches 1.

\section{Discussions}\label{sec:application}
Our work takes an important step in understanding a relatively unexplored class of
networks: the class of ad hoc networks with significant rates of addition {\em and}
deletion of nodes. To the best of our knowledge, we provide the first empirical study on
the nature of deletion dynamics in complex networks.  Using longitudinal Web crawl data
that spanned the period of one year, we discovered the preferential survival mechanism and
quantified its parameter. In order to study the implication of the preferential survival
dynamic, we analyzed a stochastic model that incorporated the standard preferential
attachment mechanism with preferential survival and showed that the power law exponent is
preserved even in the face of extremely high rate of node deletion.

As large scale network systems play an increasingly important role in our daily lives, the
dynamics identified in this work could shed light on the empirical observation of
real-world networks, and could make good candidates to be harnessed to engineer network
applications. For example, from the perspective of modeling real-world networks, the
empirical observation of PS (with $\alpha \approx 1$) and a weak DPA mechanism in the
crawled web data, can by itself explain the resilience of the PL exponent observed in the
web networks, even though the deletion rates are quite high (See
Section~\ref{ssec:model_psdpa} for the analysis results). As noted in the introduction,
however, a more complete modeling of the web data should take into consideration the
underlying fitness distribution of the pages, and such a comprehensive modeling effort is
beyond the scope of this paper.

The models studied in this paper could also find applications in the design of engineered
networks. For example, in order to develop scalable search algorithms for large scale
peer-to-peer (P2P) networks such as Gnutella, researchers have proposed efficient search
protocols that harnessed and exploited the network's power law degree distribution to
deliver search hits at a traffic cost that scales sublinearly with network size
\cite{adamic_pl,sarshar:search}. Given the unreliable and ad hoc nature of the nodes in
peer-to-peer networks, it is important to develop distributed and local protocols that
will guarantee the maintenance of the network's power law topology even in the face of
extremely high rate of node turnovers. One of the solutions proposed in \cite{sarshar} is
to introduce a compensatory mechanism where existing nodes compensate for lost edges. In
this work, however, we showed that preferential survival (PS) mechanism can stabilize the
power law exponent.   One potential way to implement PS in a P2P setup would be to enforce
an incentive mechanism to encourage high-degree nodes to remain in the network.  For
example, peers can be rewarded with virtual monetary payment in an incremental manner for
extending its availability in the network; in return, the virtual earnings can then be
exchanged for services such as priority in downloading files.  The exact reward function
can be tuned to generate a preferential survival mechanism with a deletion kernel, $D(k)
\propto k^{-\alpha}$, with $\alpha \geq 1$.  The precise implementation of the incentive
mechanism is beyond the scope of the current work and is left for future investigations.

%\section{Conclusion} %Sec VI 100 words

\section*{APPENDIX: FURTHER EMPIRICAL ANALYSIS OF WEB DYNAMICS}%\label{sec:further}
\begin{figure}[htp]
\centering
\includegraphics[height=3.0in,angle=-90]{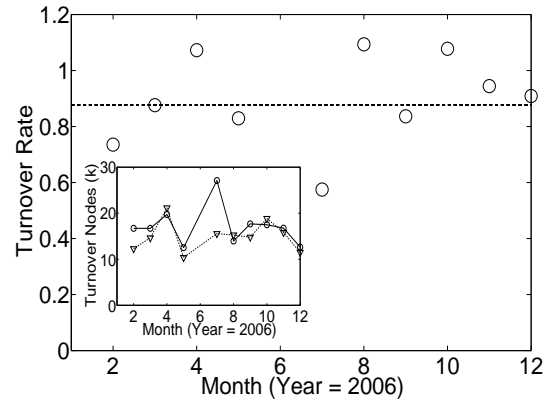}
\caption{The monthly turnover rate of the sampled Web (SET1), comprising
more than 170,000 pages, for the year 2006. The dashed horizontal line denotes the time
average turnover rate: $\bar{c}=0.88$. The inset figure shows the number of webpages (in
thousands) inserted (circles) and removed (triangles) for each month.
More than 10\% of the webpages are involved in turnovers for most months.}
\label{fig_1}
\end{figure}

{\bf Estimating the Deletion Rate.}
From SET1, we found that the deletion rate is quite high $c=0.88$ (Fig. \ref{fig_1}).
%that is, for every webpage that is added to the Web, $0.88$ webpage is removed on average.
We further found that as much as more than 10\% of the nodes are involved in turnovers (inset Fig. \ref{fig_1}).
SET2 yields very similar results indicating high rate of turnover with the average turnover rate found to be $c=0.96$.
However, these figures are overestimates since we are taking measurement from a fixed set of web hosts.

Consider the liberal assumption that the Web grows at a rate of about 35\% annually (i.e. 3\% monthly).
This assumption implies that the Web will double its current enormous size of more than 11 billion nodes \cite{gulli}
in just a little over two years.  This rough estimate is obtained by noting that the number of web hosts
has been growing at a rate of 25\% for the past few years as measured by the Netcraft server survey
\footnote{http://news.netcraft.com/archives/web\_server\_survey.html}.

Given that the Web grows at a rate of 3\% monthly and the finding that around 10\% of the webpages are removed in a month's time,
a set of new nodes with a size equal to 13\% of the network size must be inserted to achieve the 3\% monthly growth.
These figures translate to a deletion rate of $c = 10\%/13\% = 0.77$ on the Web.
Note that measurements from the literature indicates up to 20\% of webpages are deleted in a month's time \cite{ntoulas},
which will imply a even higher deletion rate.  In addition, even if we assume an unlikely
annual growth rate of 100\%, the deletion rate is still well above 0.5, which is quite significant.
%Thus, with such high rate of node turnovers, modeling the Web as an {\em ad hoc} network is appropriate.

\begin{figure*}[htp]
\subfigure[\small ]{
\centering
\includegraphics[width=2.1in]{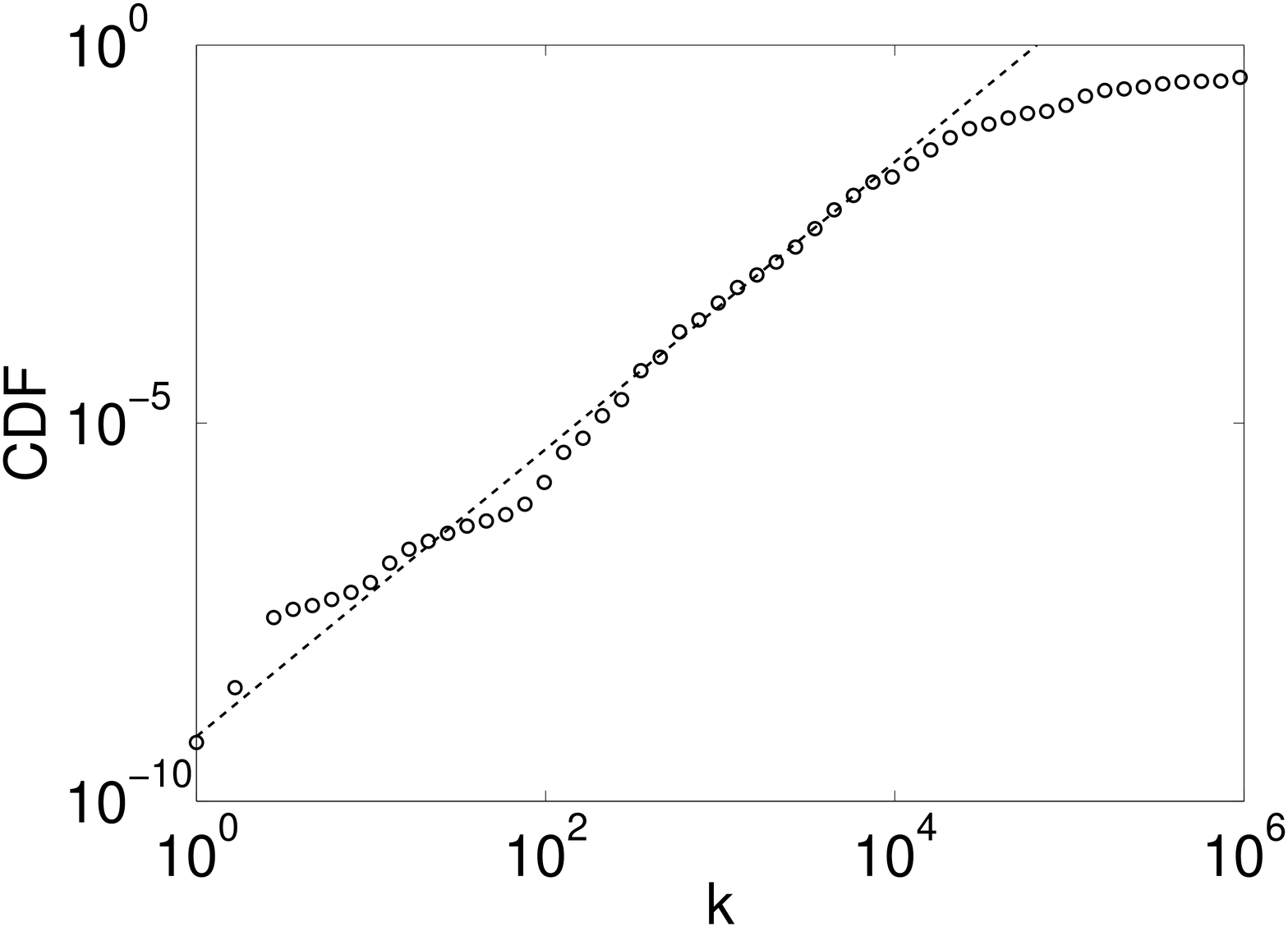}
\label{supfig_6}
}
\subfigure[\small ]{
\centering
\includegraphics[width=2.1in]{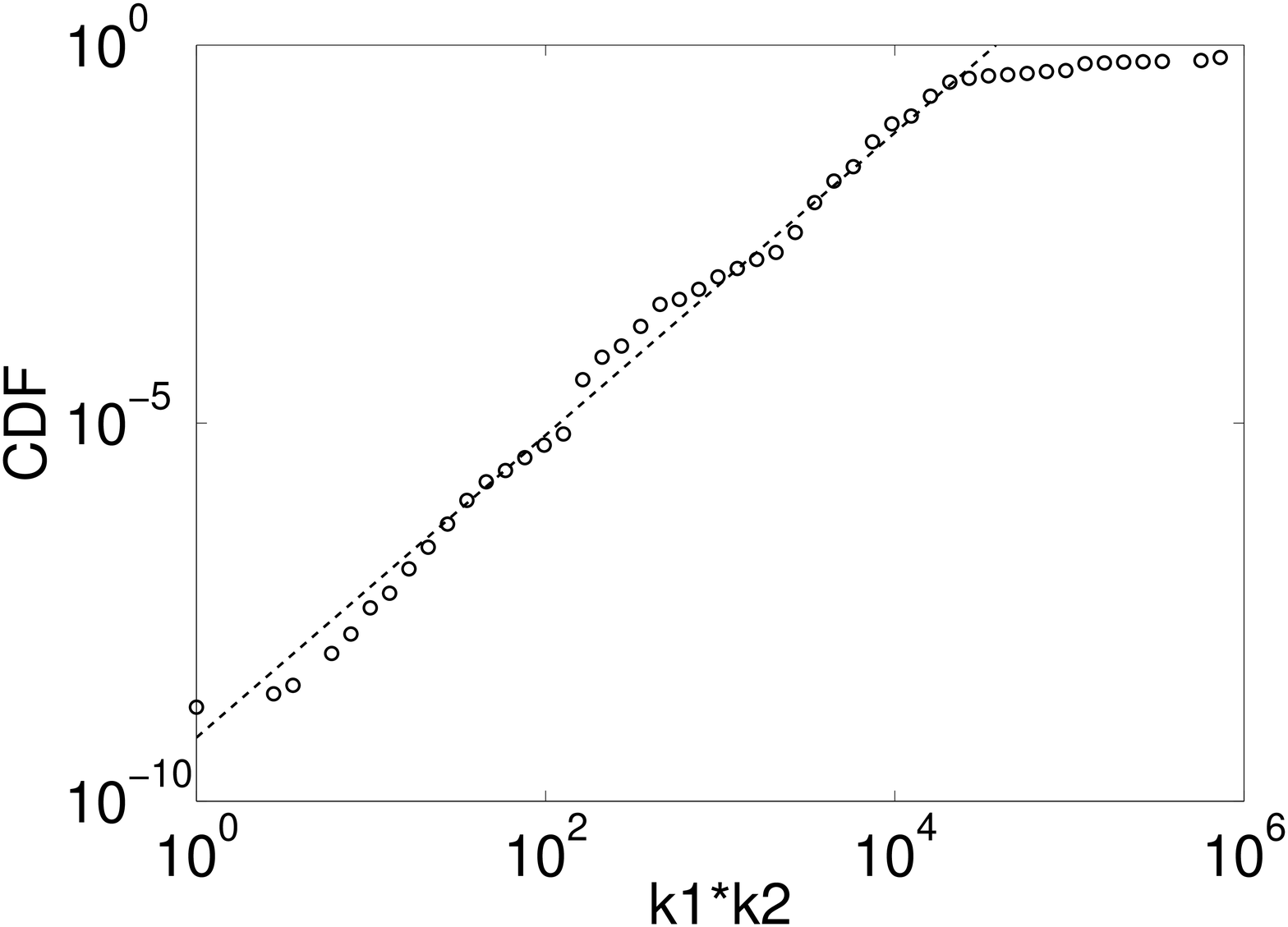}
\label{supfig_7}
}
\caption{\small (a) Empirical evidence of the preferential attachment of new webpages introduced in February, 2006.  The dotted line has a slope of
1.9 on a log-log scale in a cumulative function plot, which suggests that the preferential attachment kernel is of the form
$k^{0.9}$. The exponent of $0.9$ is very close to the exponent of $1.0$ from a linear preferential attachment kernel (also the
attachment kernel obtained by the "copying" mechanism).
(b) The figure shows empirical evidence of double preferential attachment (data from April, 2006).
Since the cumulative function is plotted, a slope of 2 (dotted line) on a log-log scale corresponds to double preferential attachment. }
\end{figure*}

{\bf Weak Mechanism of Double Preferential Attachment Edges.}  %\label{ssec:empirical_dpa}
On the Web, existing webpages often make new links to each other, where these new links
are found to attach in a double preferential manner (as determined from our empirical dataset discussed in
the next subsection). Under the assumption of
uniform deletion, we found that even with DPA, the power law exponent behaves only as
$\gamma = 1 + \frac{2(1+b)}{1-c+2b}$ (derived in Sec. \ref{ssec:model_dpa}), where $b$ is the ratio of the number of DPA edges
and preferential attachment (PA) edges (from a joining node) per time step.
Thus to get a $\gamma<3$ for $c\approx 1$, $b$ has to be $>1$; however, our
empirical data (both SET1 and SET2) indicates that the DPA edges are only a small fraction of the PA edges
(i.e., $b\approx 0.1$), and thus DPA by itself cannot explain the low power law exponent
we observe under high turnover rate (e.g. for $c=0.88$ and $b=0.1$, the predicted exponent
is $\gamma=7.9$). In the case of \emph{both} preferential survival and DPA, we use the
empirically obtained values of $\alpha=1$ and $b=0.1$, and found that the power
law exponent actually \emph{decreases} as the network experiences {\em higher} turnover
rate (see Sec. \ref{ssec:model_psdpa}). Thus, when used in conjunction with the preferential survival
dynamic, even a weak DPA mechanism (i.e. $b=0.1$) is critical in driving the power law
exponent close to $2$ in the face of extremely high rate of turnovers.

{\bf Measuring the Preferential Attachment Kernel.}
Although the preferential attachment (PA) model and the copying model \cite{barabasi,kleinberg} are widely
accepted as models of the Web, relatively few direct measurement studies \cite{jeong,newman:pa,wiki}
has been performed to validate the {\em linear preferential attachment kernel} generated by these models.
When a new node joins the existing network and attaches edges to existing nodes preferentially,
we obtain the following attachment kernel: $\Pi(k) \propto k$, where $k$ denotes the degree of the target node.
Using our data set, we performed measurements on sets of new nodes that appear every month,
and confirmed the validity of the PA hypothesis (Fig. \ref{supfig_6}). Similarly,
when a new edge emerges and attaches to two existing nodes preferentially, we obtain the following
double preferential attachment (DPA) kernel: $\Pi(k_1,k_2) \propto k_1k_2$, where $k_1$ and $k_2$ denote the degrees of the target nodes.
We repeat the same measurement on the set of new edges
that attach to existing nodes in a given month, and confirmed the validity of the DPA
hypothesis (Fig. \ref{supfig_7}).

\bibliography{del_net}

\end{document}